\documentclass[conference]{IEEEtran}
\IEEEoverridecommandlockouts
\usepackage{amsmath,amssymb,amsfonts}

\usepackage{algorithmic}
\usepackage{graphicx}
\usepackage{textcomp}
\usepackage{xcolor}
\usepackage[
backend=biber,
style=ieee,
]{biblatex}
\begin{document}

\title{Mobility Aware Edge Computing Segmentation Towards Localized Orchestration\\
}

\author{
\IEEEauthorblockN{Sam Aleyadeh, Abdallah Moubayed, and Abdallah Shami} \\
\IEEEauthorblockA{ Western University, London, Ontario, Canada\\ e-mails: \{saleyade, amoubaye, abdallah.shami\}@uwo.ca\\
}
}

\IEEEoverridecommandlockouts
\IEEEpubid{\makebox[\columnwidth]{978-0-7381-1316-6/21/\$31.00~\copyright2021 IEEE \hfill} \hspace{\columnsep}\makebox[\columnwidth]{ }}
\maketitle
\IEEEpubidadjcol
\begin{abstract}
The current trend in end-user device's advancements in computing and communication capabilities makes edge computing an attractive solution to pave the way for the coveted ultra-low latency services. The success of the edge computing networking paradigm depends on the proper orchestration of the edge servers. Several Edge applications and services are intolerant to latency, especially in 5G and beyond networks, such as intelligent video surveillance, E-health, Internet of Vehicles, and augmented reality applications. The edge devices underwent rapid growth in both capabilities and size to cope with the service demands. Orchestrating it on the cloud was a prominent trend during the past decade. However, the increasing number of edge devices poses a significant burden on the orchestration delay. In addition to the growth in edge devices, the high mobility of users renders traditional orchestration schemes impractical for contemporary edge networks. Proper segmentation of the edge space becomes necessary to adapt these schemes to address these challenges. In this paper, we introduce a segmentation technique employing lax clustering and segregated mobility-based clustering. We then apply latency mapping to these clusters. The proposed scheme's main objective is to create subspaces (segments) that enable light and efficient edge orchestration by reducing the processing time and the core cloud communication overhead. A bench-marking simulation is conducted with the results showing decreased mobility-related failures and reduced orchestration delay.
\end{abstract}

\begin{IEEEkeywords}
Edge Computing, Orchestration, Segmentation, Mobility, Clustering
\end{IEEEkeywords}

\section{Introduction}
Edge Computing was introduced to allow data processing in its locality to bypass the network backbone overhead \cite{MoubEMC}. The initial design relied on dedicated servers placed near the end-users. The design was short-lived as the number of users and services employing the design grew too quickly. The global mobile data is projected to grow 7-fold, with annual traffic almost hitting one zettabyte, according to Cisco's most recent annual report for the period (2018-2023) \cite{Intro1}. This led to the development of network virtualization and 5G networks \cite{HHVirt}. Their development helped the edge computing paradigm adapt to expected growth by allowing the services to run seamlessly on existing edge devices and idling networking infrastructure \cite{Intro2}. Hence, edge computing grew from a minor portion of the entire network into a significant part of it that is no longer a homogeneous set of servers, such as those found in core clouds. Proper management of the edge computing environment became a hot research problem \cite{MoubEdge2, ShaEdge}. The concept of dynamically shifting services to serve users better emerged in edge computing orchestration. The orchestration of services in the edge environment considers the end-user's localization, services required, and the available edge devices and their computational capabilities. Such a complex task has been rightfully delegated to the core clouds, given the edge deployment's sporadic nature makes it challenging to take it on. However, with the increased adaptation of popular low latency services such as AR, E-health, and IoV, the idea of orchestrating off-site and engaging the network backbone becomes less feasible \cite{Abs2}. A new method must be introduced to shift the orchestration process to the edge to maintain its advantages required by low latency services. While becoming more powerful and numerous, Edge devices still cannot approach the core clouds' \cite{CloudSharkh} capabilities and the resources necessary to handle the orchestration process in its current monolithic format.

The edge resources are limited by design. Thus, increasing devices' resources to resolve this issue is not feasible. A possible approach to this problem would be the segmentation of the edge environment into more manageable pieces. While promising at first glance, this approach loses favor due to the need to properly handle several entities: services,  edge devices, and end-users, each with its own set of constraints: 
\begin{itemize}

\item Services: Share a common trait of requiring low latencies, but differ in their other attributes, mainly in the amount of computational resources needed and their type; between GPU processing services in infotainment and AR apps due to heavy CPU processing such as gaming, asset tracking, and autonomous ride-sharing \cite{Intro3}. 

\item Edge devices: Differ in their computational resource capabilities as well in both type and magnitude. Additionally, they hold their own unique set of constraints, such as their coverage area, energy costs, communication method, and OPEX costs. 

\item End-users: Between smart vehicles, unmanned drones, and pedestrians, the end users' traits vary greatly, such as mobility from stationary to highly mobile, communication methods such as WiFi, 5G, and Bluetooth. Their communication format, such as V2X and D2D, as well as the services each requires. 
	 
\end{itemize}

When these entities are taken into account, the process of segmenting the edge space while safeguarding each of their requirements becomes harder to grasp. Thus, while segmenting the edge environment is paramount to our goal of edge-based orchestration, a simple system is not feasible. In contrast, an advanced system will be prohibitively complex based on the available resources of typical edge devices.
A traditional approach of geographical grid-based segmentation would not suffice as it will significantly limit the optimal local solution. On the other hand, a highly granular and advanced clustering method can achieve low optimality differences between global and local optimal. However, it will likely falter when faced with high mobility users and spiral into a continuous cycle of updates between subspaces due to users exiting and entering. Finally, end-users have several communication methods within the edge space that makes a geo-based approach ill-advised. Using geographical partitioning can isolate users from a local optimal placement due to geographical distance, even if their latency is better than a physically closer edge device.

We introduce a new system to address the above challenges. The proposed segmentation system is comprised of three different modules. The first module separates high and low mobility users into two distinct layers representing pedestrian and vehicular mobility to better manage end-user mobility. The subspaces are created by clustering each mobility layer separately with different settings tailored to mobility type while allowing the newly formed layers to overlap in the coverage area. The second module virtually localizes the end-users based on their latencies instead of geolocation in a new map. Finally, the last module uses the previous two modules' inputs to resolve frequent update pitfalls and complexity limitations. It achieves this by laxly clustering the end-users while allowing more nomadic users to exist outside of the defined subspaces.  

The remainder of the paper is organized as follows. In Section \ref{related}, the background of edge computing and related works are discussed. Then, the design and the architecture of the segmentation scheme are explained in Section \ref{sys_design}. Next, in Section \ref{sim_results}, the simulation setup is summarized, and the results are evaluated. Finally, the paper concludes in Section \ref{conc} with pointers to our future work.

\section{Related Work}\label{related}

A majority of the existing literature focuses on orchestrating micro-service chains across geo-distributed mega-scale data centers. The dimensionality problem has recently attracted attention. Bouet \textit{et al.} \cite{mecgeo} offered a graph-based algorithm that takes the maximum MEC server capacity and consolidates as many users as possible at the edge into MEC clusters to maximize edge-based processing through spatial partitioning of the geographic area. Their approach considers capacity violations while allowing the cells of the same server always to be contiguous.

Guan \textit{et al.} \cite{randoclustermec} tackled MEC region dimensionality by minimizing the number of possible handovers through clustering. They relied on a randomized algorithm dividing a metropolitan area into disjointed clusters. The proposed system is capable of finding sub-optimal partitions that can achieve their set goals.

Tran \textit{et al.} \cite{stochatgeopart} focused on creating geographically compact clusters. They introduced a novel stochastic geo-aware partitioning heuristic algorithm that offers multiple solutions for different tradeoffs between cost minimization and geo-awareness.

Lyu \textit{et al.} \cite{ offload} addressed MEC scalability problems about the massive number of devices. The author focused on resolving the issue by building a framework for offloading tasks without coordination among devices. The proposed system operated on the edge devices and involved an offloading scheme to minimize the signaling overhead.

Wang \textit{et al.} \cite{ mob5} tackled user mobility by relying on the predictability of vehicles' mobility patterns and presented a 5G mid-haul design strategy that connects each Central Unit to a suitable subset of Distributed Units. The proposed system's main aim is to manage the edge computing resources required to handle peak vehicle-to-cloud V2C application load among all Central Units, showing that it can be significantly reduced.

You \textit{et al.} \cite{fancluster} proposed an iterative Coverage Efficient Clustering Algorithm. The system aims to maximize coverage efficiency under delay constraints.  For example, resolving the Flying Ad-Hoc Networks (FANETs) conflict between area coverage efficiency and delay performance by alternately optimizing cluster heads, positions, and transmit powers in each iteration.
The works above choose to focus on delay and cost as the objectives to minimize. 

The focus of our research problem is different from the problem (latency-based partitioning), the optimization objective (orchestration load and time), and constraint (solution space size).

\begin{figure*}
	\centering
	\includegraphics[width=0.75\linewidth]{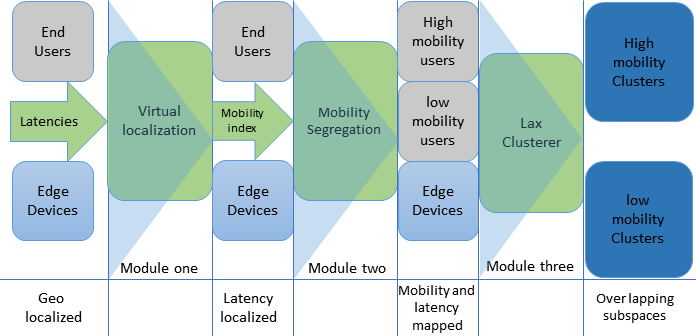}
	\caption{System overview}
	\label{fig:systemover}
\end{figure*}

\section{System Design}\label{sys_design}

The system's main aim is to segment the edge environment into smaller subspaces to allow for edge-based orchestration. However, this approach can lead to lower local optimal service placements if not adequately performed due to isolating suitable candidate edge devices. It can also lead to counterproductive re-segmentation cycles if subspaces created are not stable and require frequent updates. 
, the system breaks the process of segmenting the edge space through three modules\ref{fig:systemover}. The virtual localization module creates a latency-based map of the total edge space. Second, the mobility segregation module separates the end-users into separate layers based on their mobility. Finally, the lax clustering module uses simple clustering techniques to create sub-spaces for each end-user type sharing all edge devices within the coverage area. Below we outline the details of each module's inner workings.

\subsection{Virtual localization}
To address the end-users heterogeneous communication nature, we make use of the virtual localization module. It sets the preliminary solution space using the latencies of the end-users and edge devices to localize them. The results are a single-layered 2D map with all users and edge devices placed based on their networking latencies concerning each other. This makes it more manageable for the latter processing modules to generate a healthy subspace capable of achieving the system's target and maintaining it. 

\subsection{Mobility-based layer creation}
In this module, the localized end-user output gets fragmented into separate maps. Each user is polled for mobility and split into either low mobility representing pedestrians and stationary users or high mobility users representing vehicular and unmanned users. The results are two distinct maps of end-user overlaid upon the same edge devices map previously generated. This module’s output achieves two goals. First, segregating based on the mobility of the users further segments the edge computing environment into smaller subspaces. Second, the generated subspaces can be clustered in the following module in such a way to match their mobility better, avoiding failure conditions.

\subsection{Lax clustering}
This module uses clustering techniques to split the two maps generated by the second module into smaller subspaces. The main clustering feature is its disjoint nature, specifically the allowance of large gaps between each cluster. While the system is compatible with most clustering techniques, we choose k-means and radial clustering due to their low computational requirements when the clustering space is limited. 
The clustering of low mobility users relies on k-means due to their ability to isolate densely populated regions. At the same time, its shape will have little impact due to the users' aforementioned low mobility.
On the other hand, radial clustering is set up with additional padding to accommodate users’ mobility without allowing the subspace to deteriorate quickly. Furthermore, the clustering of low mobility users separately allows this subspace to accommodate more entities without growing too complex.

\section{Simulation and results}\label{sim_results}

To properly test our environment, a robust edge computing simulator is necessary. We considered several popular edge environment simulators such as iFogSim \cite{iFogSim}, MyiFogSim \cite{MyiFogSim}, EdgeCloudSim \cite{Edgesim}, and YAFS \cite{YAFS}. While the other simulators were more user-friendly, including a GUI interface in iFogSim, EdgeCloudSim was chosen. This choice was based on several criteria that suited our needs. The simulator is required  to have :
\begin{itemize}

 \item an easily customizable mobility and localization class to represent the end-user behavior.
 
 \item  robust generic orchestrator. 
 
 \item  modular design to support robustness and easiness.
 
\end{itemize}

 EdgeCloudSim built-in  orchestrator is based on European Telecommunications Standards Institute (ETSI) MEC orchestration method \cite{ETSI} capable of orchestrating, offloading, and load balancing between edge devices. The simulation is implemented on a workstation consisting of an eight-core CPU, 11GB DDR5 VRAM GPU, and 32 GB Ram.

The performance evaluation metrics used in simulation to evaluate  
the system performance for  minimizing the solution space without impacting the optimal placements. Additionally, ensuring the system's robustness to user mobility, especially concerning its ability to avoid the need for frequent updates are listed below:

\begin{itemize}

	\item Network delay: Corresponding to the impact of the system's segmentation on the orchestration, local versus global optimal.

	\item Task failures due to mobility: Corresponding to the impact of limiting the orchestration space and the possibility of losing end-users during the operation post orchestration.
	
	\item Task failures due to VM capacity: Corresponding to the impact of limiting the orchestration space and the possibility of losing service during the operation post orchestration due to saturating the limited set of edge devices.
	
	\item Cluster health: Corresponding to the rate of change within the segmented subspace over the run-time.
	
\end{itemize} 

To better represent the varied nature of services on the edge computing environment, the simulator implemented four service types: AR, E-health, Gaming, and Infotainment. The services were chosen to cover the unique aspects of the popular services found in the edge. Each has a unique combination of 13 attributes, including task length, active, idle period, and delay sensitivity.

The simulation was conducted using a single simulation setup but an incremental increase in the mobile device count in increment steps of a hundred over six cycles. The test was iterated 25 times for accuracy. The results below show the simulation's aggregated outcomes for the unsegmented monolithic approach, single layer (no mobility segregation), and the proposed dual-layer clustering schemes.

\begin{figure}
	\centering
	\includegraphics[width=1\linewidth]{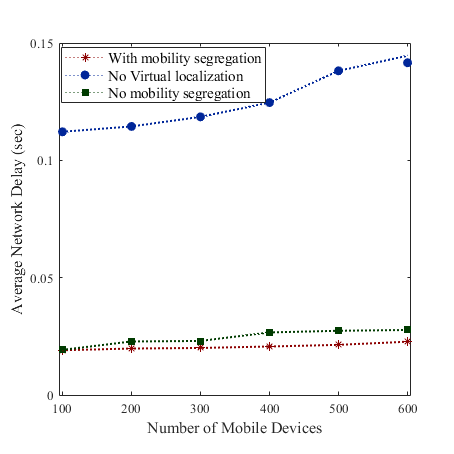}
	\caption{Impact of virtual localization on delay}
	\label{fig:localization}
\end{figure}

Fig. \ref{fig:localization} highlights the importance of the virtual localization layer on the proposed scheme's performance. Without it, the network delay remains susceptible to growth with the increase in mobile devices. However, using network virtualization, both with and without mobility segregation, led to a significant reduction in the delay end-users experience. This is attributed to promoting the latencies mapping over the geographical mapping, thus avoiding cases where heterogeneous communication methods not matching physical closeness.

\begin{figure}
	\centering
	\includegraphics[width=1\linewidth]{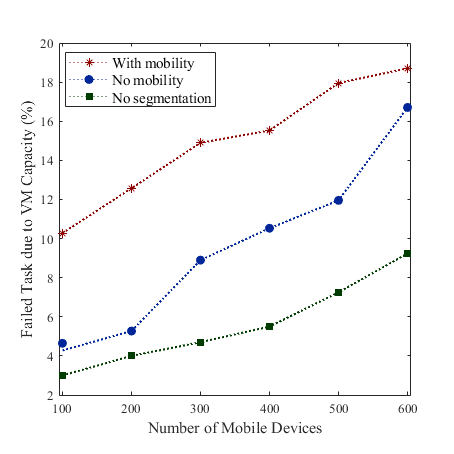}
	\caption{Impact of segmentation resource restrictions}
	\label{fig:vmimpact}
\end{figure}

In Fig. \ref{fig:vmimpact}, the scheme's performance in operating using the isolated edge resources within each subspace is illustrated. The results show that the scheme has suffered a slight setback at the lower ranges of user density. This is due to the subspace's over-saturation, causing the orchestrator to overwhelm the limited number of edge devices. The failure rate gap fluctuated within a range of approximately 7\%. Upon approaching 600 users, the difference in failed tasks is much lower. We can address these issues by introducing more complex clusterers on the edge device layer; however, segmentation reaching such high values is rare.

\begin{figure}
	\centering
	\includegraphics[width=1\linewidth]{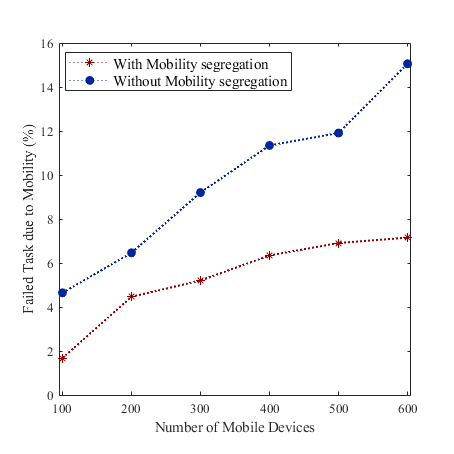}
	\caption{Impact of user mobility on edge space robustness}
	\label{fig:mobilityimpact}
\end{figure}
The segmentation process allows for the orchestration within the edge space. However, the orchestration's limited operation while being the target can become counterproductive. A great indicator of this is the failure rate due to users exiting the subspaces.

Fig. \ref{fig:mobilityimpact} shows that the system achieves low task failure relating to the user's mobility. We can observe a growing gap caused by the mobility segregation module. It plateaued at a low failure rate up to $10\%$ lower than the latter even when the number of users in the simulation approaches $600$. This is due to the tailored treatment of mobility-aware segmentation, such as the use of radial clusters with padding to prolong the period where a subspace maintains a user's relation at the cost of a marginal increase in the subspace members.

\begin{figure}
	\centering
	\includegraphics[width=1\linewidth]{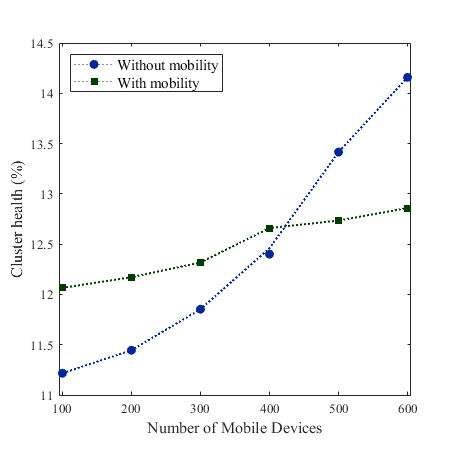}
	\caption{Impact of user mobility on cluster health degradation}
	\label{fig:clusterhealth}
\end{figure}

The created subspaces become an isolated orchestration problem. The duration of the subspace's viability must be considered to ensure that the subspaces created remain representative and do not require frequent updates. Fig. \ref{fig:clusterhealth} shows the dual-layer clustering and single-layer cluster's health degradation. The cluster health metric is calculated based on the number of clusters users lost and new users added over the total number of subspace users. This facilitates monitoring  how well each subspace segmentation can maintain its original setup after several mobility cycles. The single-layer clustering (no mobility segregation) offers marginally improved cluster health compared to its counterpart up to the $400$ mobile devices simulated. However, this advantage quickly dissipates beyond that boundary with the loss of health of more than $2\%$ compared to dual-layer clusters. This is attributed to the Lax clustering approach limiting the cluster's size based on density which remains low up to the $300$ mobile device threshold, making it easy for low mobility users to exit the subspace. However, once the users number increases beyond the 300 range, the dense regions grow in number and coverage, making it more accommodating to the user's mobility.

\begin{figure}
	\centering
	\includegraphics[width=1\linewidth]{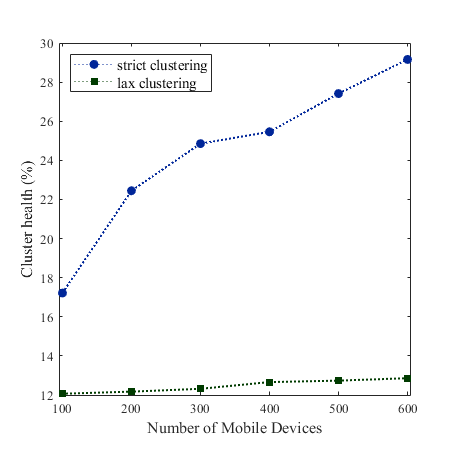}
	\caption{Impact of clustering method on subspace robustness}
	\label{fig:clusterhealthstrict}
\end{figure}

Fig. \ref{fig:clusterhealthstrict} focuses on the clustering approach and its impact on subspace robustness. With strict clustering, the system fully segments the edge space, and no user can exist outside of a cluster, while a wanted trait, as shown in the figure, leads to a significant degradation in the subspaces. The significance of lax orchestration is highlighted by its stability regardless of the increase in the number of users. Furthermore, showing the limited changes to subspaces shows that each created subspace can remain usable for much longer periods of time, thus engaging the segmentation system less often.

\section{Conclusion}\label{conc}

We proposed a system for edge computing space segmentation. The target of this system is to break down the monolithic edge environment into robust edge orchestration-friendly subspaces. To our knowledge, the approach of segregating users' mobility and virtual localization was not addressed in the edge computing current literature. We have proposed and evaluated a three-layered system. The solution, while not optimal, remains a practical approach. For future work, our next step is to refine the clustering methods concerning mobility, specifically supporting heterogeneous users mobility  types. We also aim to consider the edge device layer in our system design besides the user layer.

\printbibliography

\end{document}